\documentclass[aps, pra, twocolumn, superscriptaddress]{revtex4-2}
\usepackage{graphicx}
\usepackage{xcolor}
\usepackage{amsmath}
\usepackage{dsfont}
\usepackage{physics}
\usepackage{hyperref}
\hypersetup{
  colorlinks=true,
  linkcolor=blue,
  citecolor=blue,
  urlcolor=blue
}
\usepackage{tikz}
\usepackage{xspace}
\usetikzlibrary{fit, positioning, quotes, backgrounds}

\newcommand\inlinecomment[1]{}
\newcommand\refsec[1]{Section~\ref{#1}}
\newcommand\reffig[1]{Figure~(\ref{#1})}
\newcommand\refeq[1]{Equation~(\ref{#1})}
\newcommand\reinforce{Reinforce\xspace}
\newcommand\multipathgreedy{routed swap-ASAP\xspace}
\newcommand\swapASAP{swap-ASAP\xspace}
\newcommand\work{work\xspace} % if we are lucky, this will become "Letter," if not then just "Article"

\tikzset{
  memory/.style={
    circle,
    draw=black,
    line width=1pt,
    minimum size=2mm,
    fill=blue
  },
  station/.style={
    draw=black,
    line width=1pt,
    fill=gray,
    rounded corners,
    inner sep=2.5mm
  },
  y=0.87cm
}
\newcommand\memory[2]{
  \node[memory] #1 at #2 {};
}
\newcommand\station[2]{
  \begin{scope}[on background layer]
  \node[
    station,
    fit=#1,
    label=#2
  ] {};
  \end{scope}
}

\newcommand\fiber[3]{
  \draw[dashed, line width=1.5pt] (#1) -- node[midway, above, sloped] {$#3$} (#2);
}
\newcommand\swapedge[3]{
  \draw[line width=1.5pt] (#1) -- node[midway, above, sloped] {$#3$} (#2);
}
\newcommand\belledge[4]{
  \draw[purple, line width=1.5pt] (#1) to[#4] node[midway, above, sloped] {$#3$} (#2);
}
\newcommand{\networklegend}[1]{
\node[
    draw, 
    anchor=north west, 
    fill=white,
    nodes={inner sep=0pt},
    column sep=5pt,
    row sep=3pt
  ] at #1 {
    \begin{tabular}{cl}
      \tikz\draw[black, dashed, line width=2pt] (0,0) -- (0.1,0); & Fiber \\
      \tikz\draw[black, line width=2pt]         (0,0) -- (0.1,0); & Station \\
      \tikz\draw[purple, line width=2pt]        (0,0) -- (0.1,0); & Bell  \\
    \end{tabular}
  };
}

\begin{document}

\title{Reinforcement Learning designs Quantum Repeater Protocols on Networks beyond the Chain Scenario}

\author{Andrés Agustí Casado}
\affiliation{Center for Computational Simulation, Universidad Politécnica de Madrid}
\affiliation{Department of Theoretical Physics, Universidad Autónoma de Madrid}

\author{Álvaro Troyano Olivas}
\affiliation{Center for Computational Simulation, Universidad Politécnica de Madrid}
\affiliation{Munich Research Center, Huawei Technologies Duesseldorf GmbH, Munich, Germany}
\author{Javier Faba García}
\affiliation{Center for Computational Simulation, Universidad Politécnica de Madrid}
\affiliation{DATSI of the ETSI de Ingenieros Informáticos, Universidad Politécnica de Madrid}

\author{Luis M. Robledo}
\affiliation{Department of Theoretical Physics, Universidad Autónoma de Madrid}

\author{Vicente Martín Ayuso}
\affiliation{Center for Computational Simulation, Universidad Politécnica de Madrid}
\affiliation{DLSIS of the ETSI de Ingenieros Informáticos, Universidad Politécnica de Madrid}

\author{Laura Ortiz}
\affiliation{Center for Computational Simulation, Universidad Politécnica de Madrid}
\affiliation{DATSI of the ETSI de Ingenieros Informáticos, Universidad Politécnica de Madrid}

\begin{abstract}
Small-scale quantum repeater protocols have been shown to distribute entanglement over existing telecommunication infrastructure by means of short range quantum communications and long range classical messages. For larger networks, there are a number of theoretical proposals about how to operate the network to achieve a good performance under varying metrics. Moreover, since the dimension of the space of possible protocols grows exponentially with network size, plenty of those proposals are based on heuristics that reduce the search space. One of these is entanglement routing, which simplifies the problem on non-linear networks by allocating a series of disjoint paths and running chain protocols simultaneously on them. However, in this work we claim routing leads to suboptimal protocols. In fact, inspired by works on linear chains, we apply reinforcement learning to non-linear networks and optimize over the larger non-routing protocol space. Then, we find protocols that outperform a routed extension of the swap-ASAP protocol, even in the parameter regimes where the latter is widely recognized as the state of the art. These results indicate that routing must be discarded in favor of non-routing methods, while they pave the way towards the design of a sensible Quantum Internet architecture.
\end{abstract}

\maketitle

\section{Introduction}
\label{introduction}

First generation quantum repeaters were proposed to fight losses when distributing long-range entanglement by splitting that distance with repeaters, generating short-range entanglement between adjacent repeaters, and then perform local measurements and long-range classical communication \cite{briegel_quantum_1998}, an operation known as entanglement swapping \cite{zukowski_event_1993}. In recent years, multiple experiments have implemented this idea and consumed the final entanglement to perform characteristic tasks, such as teleportation \cite{thomas_quantum_2024} and key distribution along an untrusted repeater \cite{langenfeld_quantum_2021}, to name a few. Meanwhile, prepare-and-measure protocols are in a more mature state, sharing production networks with commercial traffic \cite{Martin2024}.
Comparing these two technologies, it stands to reason that future experiments will consider larger networks for entanglement distribution as well, which has led to the proposal of a road map towards the Quantum Internet \cite{wehner2018}. In this work we focus on one of its challenges, which is to find efficient ways to manage quantum networks beyond the chain scenario such that we maximize the entanglement rate established between two given points. That is, our objective is to find efficient repeater protocols in non-chain networks.

Previous work has typically considered long chains of repeaters and their asymptotic performance under many different metrics, for a review see \cite{Azuma_2023}. This has resulted in the proposal of multiple repeater protocols, such as tree-like arrangements of entanglement swapping rounds \cite{briegel_quantum_1998, sangouard_quantum_2011} or protocols that have repeaters swap short-range entanglement as soon as possible (swap-ASAP) \cite{kamin_exact_2023}.

Some other protocols are the result of applying machine learning methods \cite{innesta2023,haldar_2024} and, most notably, are shown to outperform the fixed, human designed, protocols mentioned above. We claim that, for arbitrary networks, there is good reason to believe that there are efficient protocols to be designed with the assistance of machine learning methods too. First of all, machine learning is known to deal well with the curse of dimensionality. With larger networks, chain or not, the dimension of state and operation spaces grow exponentially. This problem is often reduced by limiting the state space to pair-wise entangled states, but the operation space remains exponential. Secondly, proper performance metrics such as rates of some entanglement measure have complicated optimization landscapes over the network parameters, which rules out optimization methods tailored to specific metrics such as linear programming or flow problems. Because of these two reasons, being able to efficiently compute gradients often yields good results.

Another benefit of this machine learning approach is that it searches the full protocol space, even protocols outside of the set defined by a common heuristic, \textit{entanglement routing} \cite{Abane_2025}. This term stands for the design of protocols that allocate network paths and then distribute entanglement through each path simultaneously. This heuristic is often regarded as a very practical approximation, because it greatly reduces the search space to an addition of parallel routes each of them executing well known protocols tailored to chains. However, we claim that routing policies are not optimal when compared against arbitrary protocols because both entanglement and the fundamental operation of repeaters, swapping, need not follow network connectivity. Then, limiting the optimization space to routing protocols might exclude the optimal one from search space.

Therefore, in this work we focus on exploring with machine learning methods the high-dimensional space of non-routing quantum repeater protocols in networks beyond the chain scenario. In particular, we model networks with intersecting paths between target nodes, faulty generation and swapping operations, decoherence on its memories and, finally, inhomogeneity in all these parameters, as described in \refsec{network-model}. Then, we define the routing heuristic and extend the swap-ASAP policy to non-linear networks in \refsec{routing-swap-asap}. We briefly introduce some concepts from reinforcement learning and describe the process of training non-routing policies in \refsec{reinforcement-learning} as well as establish a benchmark based on the expected rate of logarithmic negativity. We present our results in \refsec{results}, which are the comparison of these learnt protocols against \multipathgreedy in terms of said entanglement rate, which is higher for the non-routing learnt policies even in networks where \multipathgreedy is expected to have high performance. Finally, we present our conclusions in \refsec{conclusions}.

\section{Network Model}
\label{network-model}

We model the quantum network as an undirected graph where nodes represent individual quantum memories (qubits) rather than full repeater stations. Unlike traditional network abstractions that treat entire repeater stations as graph nodes \cite{innesta2023,haldar_2024}, our framework shifts the fundamental unit of the graph to individual quantum memories and introduces a new type of edge to connect qubits within a repeater station. Modeling the network at repeater station level introduces unnecessary complexity, as repeaters are inherently a composite entity whose internal state and operations depends on the state and interactions of its constituent qubits among them and with the rest of the network. By defining individual qubits as the atomic graph nodes and a new type of edge that connects qubits within a repeater station, the network naturally simplifies into a collection of qubits that interact and share entanglement among them. 

\begin{figure}
    \begin{tikzpicture}[x=\columnwidth, y=\columnwidth]
\node[font=\Large\bfseries] (a) at (0, .65) {a.};
\memory{(A)}{(.3,.5)}
\station{(A)}{Alice}
\memory{(C1)}{(.55,.6)}
\memory{(C2)}{(.65,.6)}
\station{(C1)(C2)}{Repeater}
\memory{(B)}{(.9,.5)}
\station{(B)}{Bob}
\fiber{A}{C1}{p_{\text{gen}}}
\swapedge{C1}{C2}{p_{\text{swap}}}
\fiber{C2}{B}{p_{\text{gen}}'}

\node[font=\Large\bfseries] (b) at (0, .4) {b.};
\memory{(A)}{(.3,.25)}
\station{(A)}{Alice}
\memory{(C1)}{(.55,.35)}
\memory{(C2)}{(.65,.35)}
\station{(C1)(C2)}{Repeater}
\memory{(B)}{(.9,.25)}
\station{(B)}{Bob}
\fiber{A}{C1}{p_{\text{gen}}}
\swapedge{C1}{C2}{p_{\text{swap}}}
\fiber{C2}{B}{p_{\text{gen}}'}
\belledge{A}{C1}{w_1}{bend right}
\belledge{C2}{B}{w_2}{bend right}

\node[font=\Large\bfseries] (c) at (0, .15) {c.};
\memory{(A)}{(.3,0)}
\station{(A)}{Alice}
\memory{(C1)}{(.55,.1)}
\memory{(C2)}{(.65,.1)}
\station{(C1)(C2)}{Repeater}
\memory{(B)}{(.9,0)}
\station{(B)}{Bob}
\fiber{A}{C1}{p_{\text{gen}}}
\swapedge{C1}{C2}{p_{\text{swap}}}
\fiber{C2}{B}{p_{\text{gen}}'}
\belledge{A}{B}{w_1 w_2}{}
\networklegend{(-.05, .1)}
\end{tikzpicture}
    \caption{Schematic representation of one of the smallest networks. (a) In its initial state, three stations (grey rectangles) are arranged in a chain with one repeater between two end points. Each station is connected by a fiber edge (dashed line) to its neighbors. For each incoming fiber edge, there are two quantum memories (blue circles), each on a different station. Memories on the same station may be connected by a station edge (solid lines), meaning that a Bell Basis Measurement is possible between them. (b) Heralded entanglement generation (EG) takes place between memories connected by fiber edges. Successful EG creates faulty entangled states represented by Bell edges with an initial Werner parameter. (c) Entanglement swapping at the repeater station consumes two existing Bell edges to establish a Bell edge between the end points.}
    \label{example-network}
\end{figure}
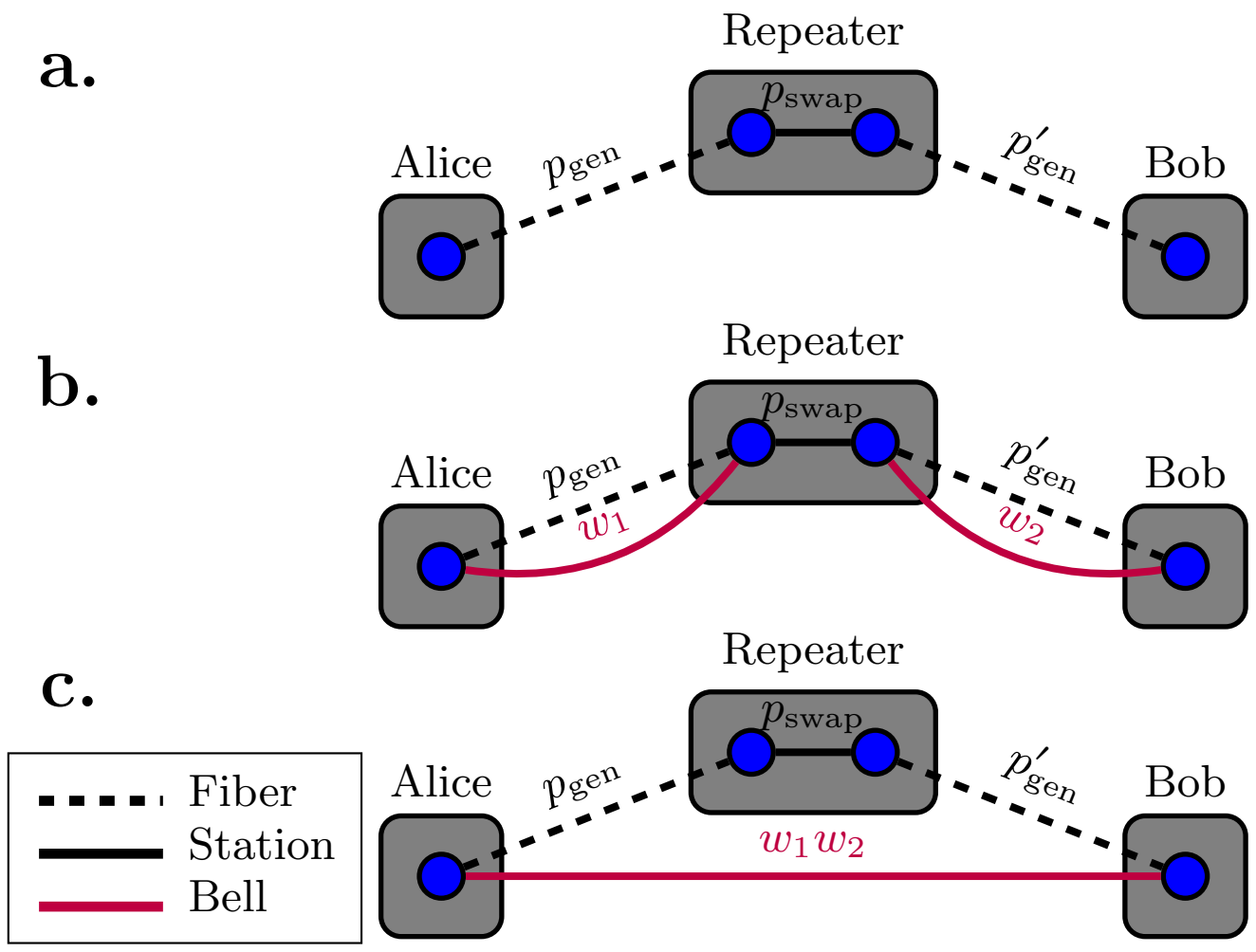

Formally, our model is a graph $G = (M, E)$ where $M$ is the set of memories and $E$ is the set of edges connecting them. To accurately reflect both the physical infrastructure and the logical state of the network, we define three different sets of edges: fiber edges $e_F \in E_F$, representing direct connections between qubits in different stations; station edges $e_S \in E_S$, linking memories within the same repeater station; and finally Bell edges $e_B \in E_B$, indicating entanglement between any two qubits. The first two kinds of edges represent the infrastructure of the network, while the last type represents its dynamical state. \reffig{example-network} offers an example of all these three edges.

The state of the network, that is, the Bell edges, changes over time due to three processes: entanglement generation between adjacent repeaters, entanglement swapping at repeaters containing two halves of entangled pairs, and decoherence. There are many different experiments showcasing each of these phenomena \cite{hermans_qubit_2022, thomas_quantum_2024, krutyanskiy_telecomwavelength_2023, cui_metropolitan_scale_2025}, but in order to draw conclusions that apply to all of them, we need to formulate a model that abstracts the details of each experiment into a common model. 

We start by introducing entanglement generation between adjacent repeaters. This operation is possible between memories $m$ and $n$ connected by a fiber edge $e_F = \{m, n\}$. Furthermore, those memories must not be entangled with any other, so there is no Bell edge $e_B \in E_B$ containing them. We choose to impose this restriction to limit ourselves to model bipartite pair-wise entanglement. Entanglement generation is a heralding operation, in the sense that it is stochastic and its success or failure are known to the network. The probability of generation is $p_{\text{gen}}(e_F)$. Conditional on its success, it generates a Bell edge $e_B$ between those two memories, that we interpret as a Werner state
\begin{align}
  \rho(e_B) = w(e_B) \ketbra{\phi^+} + (1-w(e_B))\frac{\mathds{1}}{4},
  \label{a-werner-state}
\end{align}
where $\ket{\phi^+}$ is a Bell pair and $w(e_B)$ the Werner parameter between the memories $m$ and $n$. From now on, we omit the Bell state $\ket{\phi^+}$ and only consider the Werner parameter $w(e_B)$ as a label representing the state in that edge. If, alternatively, the operation fails, no Bell edge is added to the graph. The first and second subplots of \reffig{example-network} provide a visual representation of this entanglement generation process in two different memory pairs.

The second process that has an impact on the network's dynamics is entanglement swapping. This operation is the cornerstone of quantum repeaters and allows for the generation of longer range entanglement by performing a \textit{local} measurement in two memories inside the same repeater. Here, we consider that two memories on the same station that share a station edge $e_S$ are able to perform said measurement. However, another condition must be met before performing the measurement: those two memories $\{m, n\}$ in $e_S$ must be entangled with different stations via Bell edges $e_{Bm}$ and $e_{Bn}$. Once those two conditions are met, this operation is possible and with a heralded probability of success $p_S(e_S)$. Upon success, it consumes the two Bell edges $e_{Bm}$ and $e_{Bn}$ and creates a new and longer Bell edge $e_B$ between the memories that were not $m$ nor $n$ in $e_{Bm}$ and $e_{Bn}$. See the second and third subplots of \reffig{example-network} for a representation of this process. After a short calculation, the Werner parameter of the final Bell edge is
\begin{align}
  w(e_B) = w(e_{Bm})w(e_{Bn}),
\end{align}
the product of the input Werner parameters \cite{li_efficient_2021}. Note, however, that this new Werner parameter might refer to a Werner state defined with respect a different Bell pair than the one introduced in \refeq{a-werner-state}. That Bell pair can be determined by the outcome of the measurement at the repeater, which we assume is sent to two privileged stations that are the targets to entangle, often denoted as Alice and Bob. We do not consider part of the model performing the local operations required to fix the Bell pair shared between the output memories.

The final process that has an impact on network dynamics is decoherence. As we did with entanglement generation, we abstract the details of the many different experiments performed so far and assume decoherence takes the form of the depolarizing channel, which for Werner states results in an exponential attenuation of the Werner parameter
\begin{align}
	w(e_B) \mapsto w(e_B) e^{-\frac{\Delta t}{\tau}}
  \label{decoherence}
\end{align}
where $\Delta t$ is the time step of the simulation and $\tau$ is the composite coherence time of the two memories storing this state. Finally, if the Werner parameter of a Bell edge $e_B$ drops below one third, there is no entanglement between those memories and the Bell edge is discarded from the simulation.

Now that we have introduced the three fundamental processes responsible for the network's dynamics we turn our attention to their composition over discrete time steps. Both generation and swapping act on a pair of memories connected by either a fiber or a station edge. Two operations that act on different memories commute within the formalism of quantum mechanics and therefore we assume that they can be run on the same time step. However, there is a caveat with regards to the evolution of the network when two or more copies of a swapping operation are performed simultaneously. If two repeaters are ready to swap on station edges $e_{S1}$ and $e_{S2}$, and one Bell edge is an input to both swaps, then the probability of success of the whole process is the multiplication of the two $p_S(e_{S1})$ and $p_S(e_{S2})$. On success they delete their input Bell edges and a new edge is created between the two memories outside the repeaters. But, if either of the swaps fails, a simple calculation shows that all entanglement is lost, including both input pairs of the successful repeater. Some of us denominate this composite operation a two-repeater entanglement swapping or 2-ES for short \cite{faba_performance_2026}. The generalization to $n$-ES is straightforward. Finally, note that for larger networks in the number of memories $M$, the number of possible Bell edges grows as $O(M^2)$. The number of fiber and station edges grows as $O(M)$ as long as there is an upper bound to both repeater and memory degree in the graph. However, since commuting actions can be run in parallel, creating new composite actions, the set of actions grows as $O(\exp(M))$. Moreover, those composite actions have a great number of possible final states, and they are iterated over time, leading to a large set of possible trajectories for the system. In order to apply heuristics or machine learning to find protocols that map network states to useful actions, we need to have a performance metric to act as a compass towards efficient protocols in this exponentially large space.

We introduce two distant target nodes in the network, Alice and Bob. Broadly speaking, the objective of a protocol is to produce entanglement between them, also known as end-to-end (E2E) entanglement. Previous works measure E2E entanglement with metrics that can be sorted into, at least, four groups. Some maximize the secret key rate \cite{li_efficient_2021}. We discard this approach on the grounds that it binds the protocol to a specific application. Others try to maximize the number of Bell edges created between Alice and Bob, or to minimize the time taken creating them \cite{innesta2023}. The third group notes that there are two objectives: the number/frequency of those pairs and their quality, often characterized by fidelity \cite{haldar_fast_2024}. Here, we take a joint approach an use the addition of logarithmic negativity $E_N$ over all E2E Bell edges produced over time \cite{vidal_computable_2002}. This quantity is an entanglement measure for two-qubit states. Moreover, it is an additive entanglement monotone, that is, if Alice and Bob share two pairs of memories in states $\rho_{1}$ and $\rho_{2}$ then
\begin{align}
  E_{N}(\rho_1 \otimes \rho_2) 
  = E_{N}(\rho_1) + E_{N}(\rho_2)
\end{align}
This property is crucial to our model. Because of it, protocols that produce many low quality E2E Bell edges can be compared against low-rate and high-quality protocols by adding this quantity over all the entangled pairs generated by both protocols over a common period of time and dividing by that time. This is the entanglement rate that we use in this work.

\section{Routing swap-ASAP policies}
\label{routing-swap-asap}

Given the network model, its operations and the entanglement rate defined in \refsec{network-model}, we now introduce heuristics that try to guide the design of efficient repeater policies while not using machine learning.

The first example of such a policy is \swapASAP \cite{kamin_exact_2023}, which results from the intuition that, given the state of a chain of repeaters, doing as many commuting actions as possible should lead to good results. So, if the network starts with no entanglement, the first action attempts to generate all fiber links. The second action generates again in the links that failed and, if some repeaters contain two Bell pairs, it swaps that entanglement right away. It stands to reason that this policy produces quickly E2E entanglement in the chain. 

However, it is not clear how to generalize that heuristic to networks beyond the chain scenario. If given a state one chooses to run as many commuting operations as possible, short range entanglement will be swapped indeed, creating entanglement of a longer range. But it might not be getting closer to the target nodes and will be lost in the non-trivial layout of the network. Conveniently enough, previous literature considering these kind of networks has created another heuristic, \textit{entanglement routing} \cite{Abane_2025}. This heuristic reduces the task of designing network protocols by taking the pair of target nodes and slicing the network infrastructure into a set of disjoint paths connecting those nodes. Then, chain based methods can be used on each path or route in parallel.

Here, we propose to apply this idea to the \swapASAP policy in the following way. Given a network and two target stations, there is a set of possible simple paths connecting them. We select the path that has highest multiplication of heralding probabilities along its fiber and swap edges. And then we select the second highest ranking path that is disjoint from the first, and so on until we run out of paths. Then, chain-based \swapASAP can be executed simultaneously on all selected paths. We call the resulting policy \multipathgreedy. Just as with \swapASAP, it stands to reason that this policy will often entangle the target nodes.

However, these are just heuristics. It has been shown in chains that \swapASAP is not optimal under realistic conditions \cite{innesta2023}. At least one circumstance in which \swapASAP is clearly suboptimal can be easily understood: If a repeater contains two halves of Werner states with vastly different quality, it might make sense not to swap them, but discard the worst one and generate it again. In fact, there might be other settings in which \swapASAP leads to suboptimal results. We conclude that if \swapASAP is not optimal, neither is \multipathgreedy. However, we believe that there are more reasons behind that disadvantage than memory decoherence. Therefore, in the research to follow we consider networks with high coherence times so that \swapASAP stays close to optimal in the routed paths. Then, if we manage to find better policies than \multipathgreedy in this parameter regime, we can conclude that routing is a heuristic that, although helpful at first, leads to suboptimal policies. This will be the objective of the remainder of this work.

\section{Reinforcement Learning Model}
\label{reinforcement-learning}

Deep Reinforcement Learning (DRL) can be considered the application of neural networks to control theory. It deals with the relation of two elements: an agent and an environment. In this work, the agent can be identified as a neural network with depth greater than one composed with other functions to become the quantum repeater policy, that is, a map from states to actions. Meanwhile the environment is the quantum network. It is unfortunate that the word environment is used in the study of open quantum systems too, but we point out that our network is an open system only in the sense that its memories decohere according to \refeq{decoherence} and the word environment will only be used in the DRL sense. And in DRL, the environment is characterized by states that are feed into the policy, and it evolves depending on the actions taken by said policy. DRL has the advantage of being able to approximate arbitrary policies and estimate gradient loss through efficient backpropagation, see \cite{foundations_drl} for an introductory reference.

From the many algorithms in DRL we use \reinforce \cite{williams_simple_1992}. We choose this algorithm because of both its simplicity and the fact that it is well suited for continuous state and discrete action spaces, as is the case of the model in \refsec{network-model}. In \reinforce, time is discretized in time-steps $t$. Its agent takes the current state of the environment $s_t$, which must be a vector, and applies sequentially the neural layers indexed by $l\in\{1,L\}$. On each layer, the input vector $v_l$ is multiplied by learnable weights $W_l$ and added biases $b_l$. A non-linear activation function $\sigma$, which in this work is set to the ReLU function, is then applied to each entry of the retulting vector
\begin{align}
  v_{l+1} = \sigma(W_l v_l + b_l),
  \label{activation-vectors}
\end{align}
for each layer $l$ and where $v_0$ is $s_t$. In order to identify the output of the network $v_L$ with actions, we set this vector to have as many entries as actions allows the environment. Then, we turn the entries of $v_L$ into probabilities by applying a softmask map
\begin{align}
  \pi(a|s_t) 
  = \frac{e^{-v_{La}}}{\sum_{a'} e^{-v_{La'}}},
  \label{policy}
\end{align}
then the agent creates the action $a_t$ by sampling an index from the policy vector in \refeq{policy} and sends it to the environment. It, in turns, updates its state to $s_{t+1}$ by sampling from its own transition probability distribution $p(s_{t+1}|s_t,a_t)$ given by the rules introduced in \refsec{network-model}. Notice that the composition of $\pi$ and $p$ creates a Markov decision process that might be tuned by the learnable weights $W_l$ and biases $b_l$. These transitions might result in rewards $r(s_t, a_t, s_{t+1})$ and this process is iterated until either a halt state or a maximum number of iterations are reached, resulting in an episode or trajectory $\tau$. The sum of rewards over an episode is the total return $G(\tau)$ of said episode. Finally, the objective of a RL algorithm is to update the weights $W$ of the agent in such a way that it maximizes the expected total return:
\begin{align}
  \max_{\{W_l, b_l\}_l}
  \mathds{E}_{\tau \sim \pi} [G(\tau)].
  \label{RL-optimization}
\end{align}

Now that we have introduced the basic notation of DRL, we refer the reader to \cite{foundations_drl} for implementation details on how to numerically solve \refeq{RL-optimization} by stochastic gradient descent. Here, we build on the notation introduced so far to discuss the application of this algorithm to the design of non-routing quantum repeater protocols specifically. 

As mentioned in \refsec{network-model}, there is an exponential number of actions to be run at a given time in the network. Since $\pi(a|s_t)$ in \refeq{policy} needs to have a dimension equal to the number of actions, then the last layer of the neural network has an exponential number of neurons with the quantum network size. To solve this problem, we turn to action serialization \cite{pierrot}. We consider that the space of actions is only composed of elementary actions, that is, single generations or swaps $A = E_F \cup E_S$ which grows as $O(M)$. We sort the actions in some arbitrary but constant order and then, at each reinforcement loop iteration, the agent chooses a single elementary action $a$ and is sent to the environment. The network does not evolve because of it, but rather keeps it in an action cache $A_C$. Broadly speaking, the agent now creates composite actions over multiple iterations. Once the agent is done building a composite action, it selects a new action $a_{\text{execute}}$ which tells the environment to take $A_C$, run it, evolve a single physical time-step and empty the cache.

For the state encoding into vectors, we first considered computing the adjacency matrix of the network's Bell edges. Then, we put in the non-zero entries of that matrix the Werner parameter of the edge it represents. However, in order to keep reinforcement loop iterations markovian, we need to extend the state with an encoding of the agents action cache. Intuitively, it makes sense that the agent needs to take the cache as input in order to know what actions to commit next. Therefore, we create a vector with $\#A$ entries and use the previously introduced action sorting to put ones in its entries if the action associated with that index is present in the cache. Then, we append this vector to the flattened Bell adjacency matrix and set the first layer of the neural network to match this dimension. This results in a state encoding with growth $O(M^3)$, a fair price to pay for a polynomial action space.

Another DRL technique we use is action masking \cite{hou_exploring_2023}. Given a state, the policy discussed so far may produce an output action that makes no physical sense. For instance, it may output an action that does not commute with the action cache, leading to an invalid composite action. Another example is the choice of actions we know will fail, such as swapping with no input entanglement. These situations can be avoided if the last layer of the neural network in \refeq{activation-vectors} is modified with a mask that sets the activations of impossible actions to the smallest representable floating point number. Then, applying the softmax layer in \refeq{policy} results in a policy that assigns zero probability to impossible actions. Note that this mask must be built from the information of the state encoding alone. Other actions removed by the mask are entanglement generation in links with at least an entangled memory and the $a_\text{execute}$ action on an empty cache.

Finally, we introduce the reward function that leads to the total return which DRL maximizes in \refeq{RL-optimization}. Back in \refsec{network-model} we introduced the multiple performance metrics the literature studies when analyzing quantum repeater protocols. We settled for the expected rate of E2E logarithmic negativity, so it makes sense to choose a reward function that turns total return into that rate. In order to do so, we set $r(s_t, a_t, s_{t+1})$ to the sum of the logarithmic negativities of all E2E Bell edges in $s_{t+1}$. More informative reward functions could be used, but they tend to reduce the performance of the final protocols and the serializing and masking techniques introduced above have been enough to make the model trainable.

\section{Results}
\label{results}

\begin{figure}
  \includegraphics[width=0.45\textwidth]{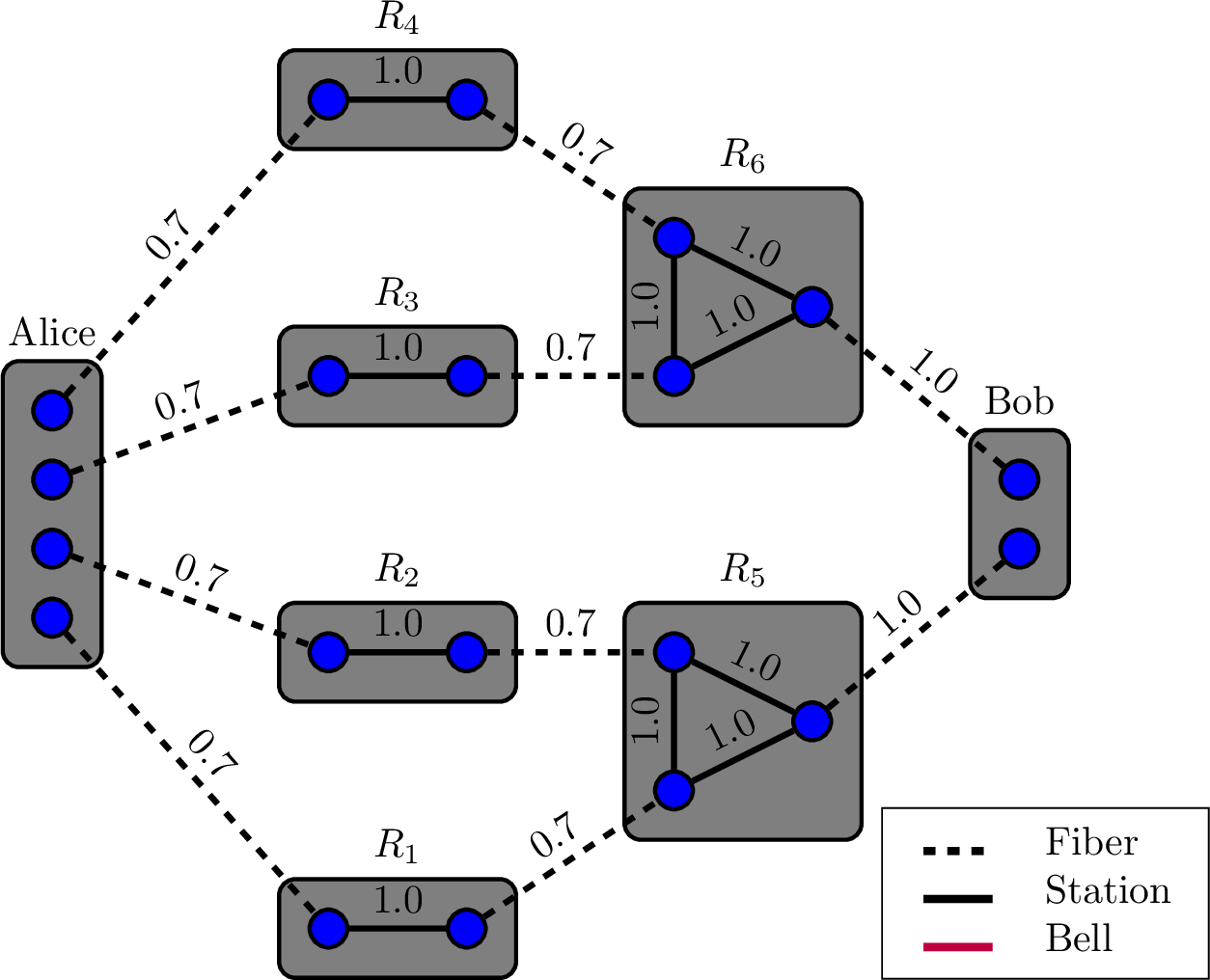}
  \caption{A typical network used to train Reinforce. For the definition of each node, edge colors and labels see \reffig{example-network}. In this network, probability of generation and swap range between 0.7 and 1. Initial Werner parameter of generated states is $1$ and the memories have no decoherence. These parameters are ideal so that we can use \multipathgreedy as a baseline with which to compare our \reinforce algorithm. Notice that between the end nodes Alice and Bob there are multiple possible paths, and plenty of them intersect with each other, a crucial property to make apparent that routing leads to suboptimal policies.}
  \label{network}
\end{figure}

In this section we introduce the results of training the \reinforce agent described in \refsec{reinforcement-learning} in multiple networks. We start with a representative network that can be found in \reffig{network}. It consists of eight stations of varying degree. Two of these stations are regarded as the target of entanglement distribution and labeled as Alice and Bob. Notably, these two stations have a very different degree: One is connected to four fibers and the other one to two. The intermediate stations have varying degree as well, between two and three. This leads to a network with multiple intersecting simple paths between Alice and Bob, which will be a crucial property when it comes to show the advantage of the \reinforce policy against routing policies. The network parameters are inhomogeneous with probabilities of heralding generation and swap varying from 0.7 to 1. The final Werner parameter after successful link generation is 1 and the memories time of coherence is infinite. Again, these are ideal parameters intended to make swap-ASAP optimal in each route. 

\begin{figure}
  \includegraphics[scale=0.6]{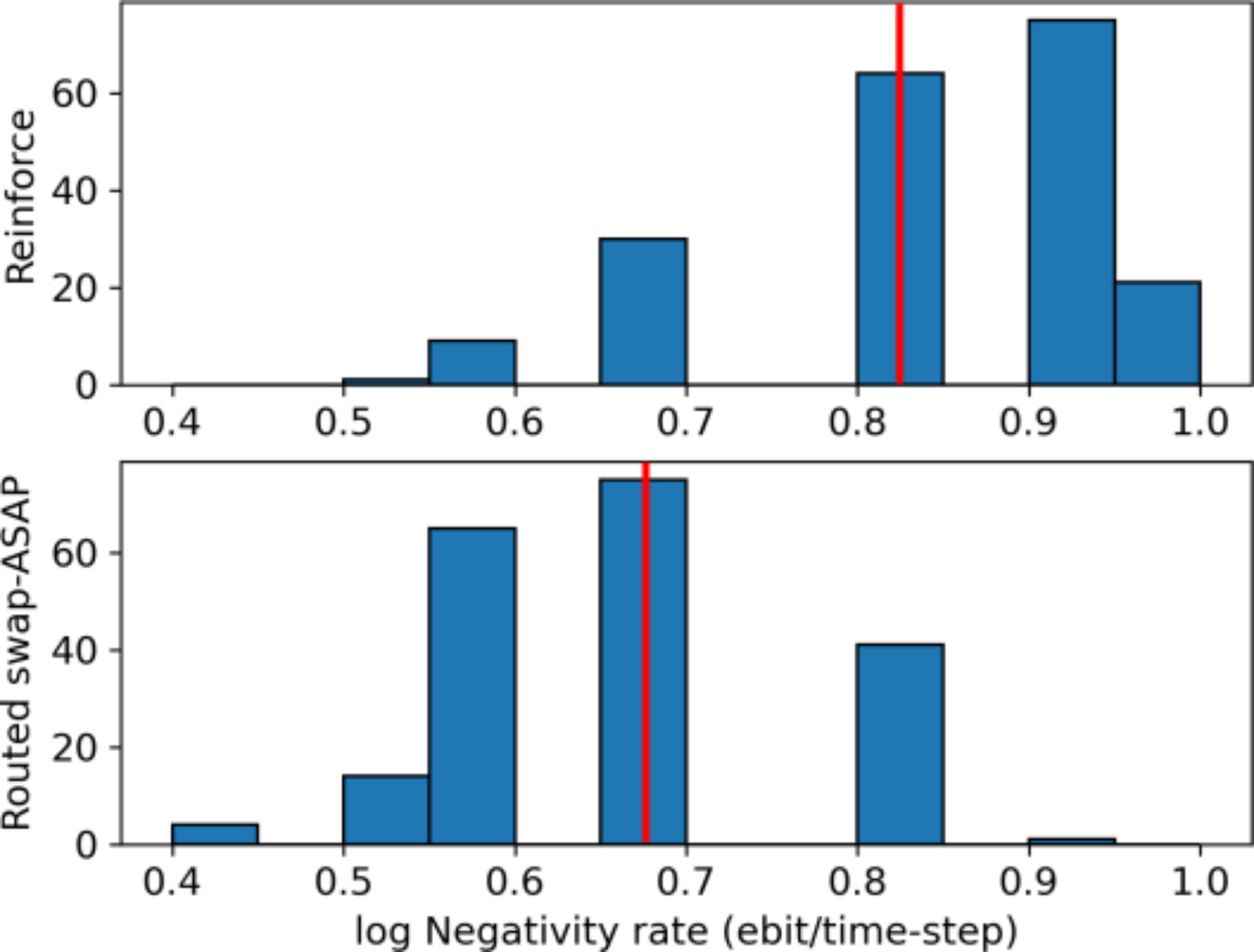}
  \caption{Histograms showing the distribution (blue) and mean (red) log negativity rates defined in \refsec{network-model} obtained in 200 simulations using either the trained \reinforce policy (upper plot) or \multipathgreedy (lower plot) in the network of \reffig{network}. Each simulation lasted 20 time steps. The log negativity rate is obtained by adding up the log negativity of each Alice-Bob memory pair over time and dividing by the simulation's duration. Non-routing Reinforce obtains a rate of $0.825 \pm 0.101$ ebits/time-step while \multipathgreedy obtains a lower rate of $0.680 \pm 0.101$ ebits/time-step, leading to the claim that the optimal protocol does not route entanglement.}
  \label{agent-performance-on-network}
\end{figure}

Notice that the layout of this network bounds entanglement rate between zero and one ebit per time step. This is due to the fact that, per time step, Bob can hold at most two perfect Bell pairs with Alice. And those two Bell pairs require at least two time steps to be created: one time step for the generation of all intermediate Bell pairs and one time steps for the $n$-body swaps. The rest of the parameters of the network only reduce this rate by introducing failures in the generation and swapping of some links for both heuristic and machine-learnt policies. Knowing the range of values for the entanglement rate, we  present the performance of both \reinforce and \multipathgreedy on \reffig{agent-performance-on-network}. The \reinforce agent is trained during 400 batches of 100 episodes of 20 time-steps each, with a learning rate of $0.0025$ using the ADAM optimizer. In the upper plot of \reffig{agent-performance-on-network} we present a histogram of the logarithmic negativity rate obtained over 200 simulations after training the agent. The obtained average entanglement rate and standard deviation are $0.825 \pm 0.101$ bits per time-step. Meanwhile, in the lower plot of \reffig{agent-performance-on-network} we present the logarithmic negativity of \multipathgreedy which has and average value and standard deviation of $0.680 \pm 0.101$ bits per time-step, well below the rate of \reinforce in units of their standard deviations. By inspecting the trajectories in the simulations, we find that \reinforce is able to adapt to both the stochastic and non-commuting nature of the elementary operations of the network while routing policies do not. That is, the network might end up by chance in a state that could lead to long range entanglement if the swaps were arranged in a path that does not commute with the best routes, the paths used in \multipathgreedy.

In order to strengthen this claim, we indicate that the results presented so far hold for other networks as well. Though, not all networks. As a consistency check, we report that both \reinforce and \multipathgreedy have the same performance in chains of repeaters containing memories with very high coherence times, as it is expected \cite{innesta2023,haldar_2024}. For the non-linear networks where we expect \reinforce to improve on \multipathgreedy, we could repeat the analysis presented in \reffig{network} and \reffig{agent-performance-on-network}. But we believe this approach adds little to this work. Instead, we use our insight obtained by inspecting the actions taken by trained agents and we introduce a function that assigns a score to a quantum network that summarizes whether we expect \reinforce to improve over \multipathgreedy. Later on, we will present the improvement in entanglement rate as a function of this score for multiple networks.

We name this quantity the accumulated non-commuting score, since it represents how many non-commuting resources a network has. It is computed as follows: Given a network and a pair of target nodes, there is a set of simple paths connecting them. These paths can be ranked based on a score which is equal to the product of all the probabilities of success in their edges. The higher the score for a path the higher the entanglement rate expected from that route. As stated in \refsec{routing-swap-asap}, \multipathgreedy takes as a route the path with best score. The second route is the next best disjoint path to the first route, and so on until we run out of disjoint paths. After this routing process ends, there are still paths that are discarded, because they intersect with the selected paths. It stands to reason that the presence of these paths is responsible for the advantage of non-routing policies. Therefore, we define the accumulated non-commuting score as the sum of the scores of all these discarded paths. In order to increase the familiarity of the reader with this score, we point out that its values range from zero to the number of discarded paths. If the score is zero it means there are no discarded paths or it is very unlikely to obtain E2E entanglement from them. If it is equal to the number of discarded paths, then all actions on those paths (and the selected routes) are ideal and a high E2E entanglement rate is expected.

\begin{figure}
  \includegraphics[scale=0.59]{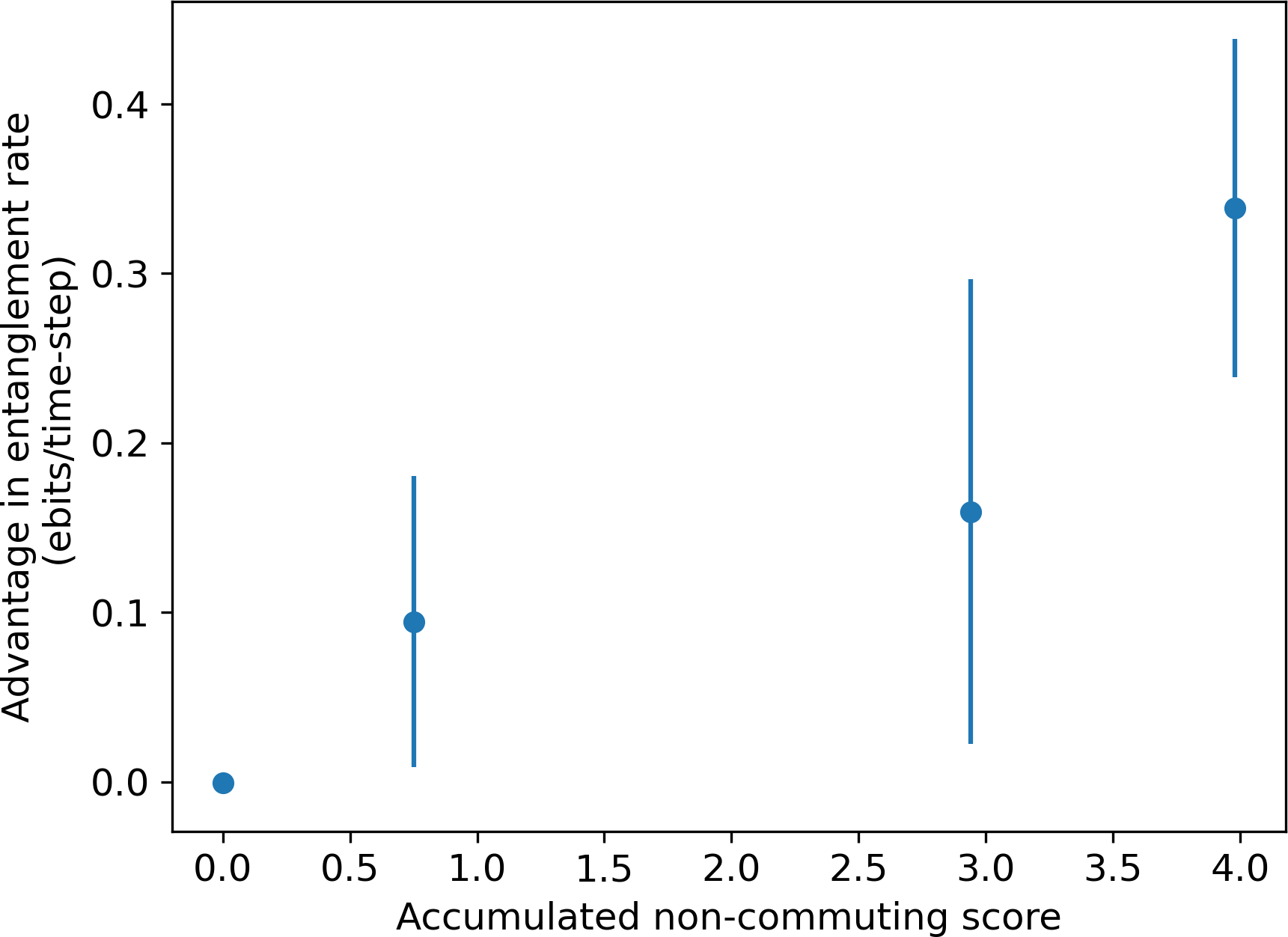}
  \caption{Logarithmic negativity rate of \reinforce minus the rate of \multipathgreedy for different networks, as a function of the proposed accumulated non-commuting score of each network. We present this figure as a summary of the main point of this article: If entanglement routing is used in a network with many intersecting routes, the resulting protocol may be outperformed by non-routing strategies, such as \reinforce. We point out that the agent trained in the network of \reffig{network} contributes the data point at score around $3$. The network with zero score is a five station chain and, as expected, the trained agent has the same performance as the routing strategy, resulting in  zero advantage. The rest of the networks require training an agent for each of them, and for information on their layouts and parameters we refer the reader to the accompanying code repository \cite{code}.}
  \label{the-plot}
\end{figure}

As our final results, we introduce \reffig{the-plot}. It represents the average increase and standard deviation of entanglement rate of \reinforce minus \multipathgreedy's as a function of this accumulated non-commuting score for multiple different networks.  These networks are the chain of length 5, the kite network, the $2 \times 3$ and the $3 \times 3$ square lattices in addition to the previously presented \reffig{network}. Their exact layouts, parameters and agents' weights are provided in the publicly available repository accompanying this work \cite{code}. We choose these networks because they are small in the number of memories while having an increasing amount of intersecting paths, that is, an increasing accumulated non-commuting score. We find that these networks and the performance of the agents trained on them confirm our insight obtained by inspecting the actions of the agent acting on the network from \reffig{network}. When the score is high, it is likely that the operation of the network will entangle its memories in ways that could lead to E2E entanglement if the next operations were not arranged along the initially best scoring routes. Then, \reinforce is able to adapt and produce that E2E pair while \multipathgreedy cannot, leading to an increase in \reinforce's entanglement rate.

\section{Conclusions}
\label{conclusions}

In this work we have introduced a network model that summarizes the three processes that are expected to be central to entanglement distribution in telecommunication networks: entanglement generation, swapping and decoherence. Moreover, all these processes are allowed to be faulty and inhomogeneous throughout the network. A brief summary of performance metrics about end-to-end (E2E) entanglement generation in these networks is presented, leading to our choice, the rate of logarithmic negativity, that we use in the remainder of this \work. Then, we introduce the reader to the exponential growth of the set of possible protocols that try to produce E2E entanglement as a function of network size and we mention that previous literature typically turns to heuristics to reduce the size of the protocol search space. Entanglement routing is a very common one, which sends the entanglement along predefined routes in the network.

Our central claim is that this heuristic does not solve the optimization problem of finding efficient quantum repeater protocols in networks beyond the chain scenario. To back this claim, we develop a reinforcement learning model tailored to the exponentially large set of non-routing protocols.  In particular, we introduce a model based on Deep Reinforcement Learning as well as an action serialization procedure, together with action and state vector encodings and masking, resulting in a polynomial model growth. Finally, we train the model on multiple networks of various types and show that \reinforce obtains a higher E2E entanglement rate than heuristic entanglement-routing protocols consistently. Additionally, we provide a function that maps a given network to a score that is shown to be correlated with the improvement of \reinforce over routing policies.

These results provide valuable information in the road map towards the Quantum Internet and the design of its architecture. Entanglement must not be regarded as a package of information to be routed along paths of the network, but a correlation among the network's elements which may not follow network connectivity. We conclude that entanglement routing should be discarded or, if used, justified with strong evidence about its convenience due to other merits than E2E entanglement rate. As part of said road map toward the Quantum Internet, this model could be extended in the following ways: Firstly, the network model could be extended to include distillation operations, which could have a fundamental impact on its performance. Secondly, the model could be extended to include classical communication delays in a more nuanced way. At present time, heralded signals are all sent to the reinforcement learning agent, which replies to each node with the actions expected from them. These delays set the minimal time step the model can use. This time step could be reduced if the agent is replaced by agents distributed over the network that work on partial information about the network \cite{haldar_reducing_2025}. Finally, the distant nodes to be entangled are fixed in this framework. However, more pairs of client nodes could be added, even in a dynamical way, with little change to the algorithm.

\section{Acknowledgements}

This work has received funding from the European Union's Horizon Europe research and innovation programme under grant agreement No 101297140 (QIA-Phase2) and No 101114043 (QSNP). A.A.C. acknowledges funding from the CAM Programa TEC-2024/COM-84 QUITEMAD-CM. The authors gratefully acknowledge the computer resources at Artemisa, funded by the European Union ERDF and Comunitat Valenciana as well as the technical support provided by the Instituto de Fisica Corpuscular, IFIC (CSIC-UV).

\bibliographystyle{apsrev4-2}
\bibliography{reinforce-ent-dist-net.bib}

\end{document}